\documentclass[aps,prl,reprint,superscriptaddress,showpacs,showkeys]{revtex4-2}

\usepackage{graphicx}    
\usepackage{amsmath}     
\usepackage{amssymb}     
\usepackage{hyperref}    
\usepackage{color}       
\usepackage{physics}     
\usepackage{siunitx}     
\usepackage{bm}          
\usepackage{gensymb}
\usepackage[utf8]{inputenc}
\usepackage[inkscapelatex=false]{svg}

\begin{document}

\title{Lifetime-Limited and Tunable Emission from Charge-Stabilized Nickel Vacancy Centers in Diamond}


%



\author{I. M. Morris}
\email{morrisia@msu.edu}
\affiliation{Michigan State University, Department of Physics and Astronomy, East Lansing, MI USA}

\author{T. L\"uhmann}
\affiliation{University of Leipzig, Felix Bloch Institute for Solid State Physics, Leipzig, Germany}

\author{K. Klink}
\affiliation{Michigan State University, Department of Physics and Astronomy, East Lansing, MI USA}

\author{L. Crooks}
\affiliation{Michigan State University, Department of Physics and Astronomy, East Lansing, MI USA}

\author{D. Hardeman}
\affiliation{Element Six Global Innovation Centre, Fermi Avenue, Harwell, Didcot, OC11 0QR, UK}

\author{D.J. Twitchen}
\affiliation{Element Six Global Innovation Centre, Fermi Avenue, Harwell, Didcot, OC11 0QR, UK}

\author{S. Pezzagna}
\affiliation{University of Leipzig, Felix Bloch Institute for Solid State Physics, Leipzig, Germany}

\author{J. Meijer}
\affiliation{University of Leipzig, Felix Bloch Institute for Solid State Physics, Leipzig, Germany}

\author{S. S. Nicley}
\affiliation{Michigan State University, Department of Physics and Astronomy, East Lansing, MI USA}
\affiliation{Coatings and Diamond Technologies Division, Center Midwest (CMW), Fraunhofer USA Inc., 1449 Engineering Research Ct., East Lansing, MI 48824, USA}

\author{J. N. Becker}
\email{becke183@msu.edu}
\affiliation{Michigan State University, Department of Physics and Astronomy, East Lansing, MI USA}
\affiliation{Coatings and Diamond Technologies Division, Center Midwest (CMW), Fraunhofer USA Inc., 1449 Engineering Research Ct., East Lansing, MI 48824, USA}


\date{\today}

\begin{abstract}
The negatively charged nickel vacancy center (NiV$^-$) in diamond is a promising spin qubit candidate with predicted inversion symmetry, large ground state spin orbit splitting to limit phonon-induced decoherence, and emission in the near-infrared. Here, we experimentally confirm the proposed geometric and electronic structure of the NiV defect via magneto-optical spectroscopy. We characterize the optical properties and find a Debye-Waller factor of 0.62. Additionally, we engineer charge state stabilized defects using electrical bias in all-diamond p-i-p junctions. We measure a vanishing static dipole moment and no spectral diffusion, characteristic of inversion symmetry. Under bias, we observe stable transitions with lifetime limited linewidths as narrow as 16\,MHz and convenient frequency tuning of the emission via a second order Stark shift. Overall, this work provides a pathway towards coherent control of the NiV$^-$ and its use as a spin qubit and contributes to a more general understanding of charge dynamics experienced by defects in diamond.    
\end{abstract}


\maketitle

Color centers in diamond, in particular the nitrogen vacancy (NV) and group-IV vacancy (SiV, SnV, GeV, PbV) complexes, are leading candidates for solid-state quantum networking, memory and computing applications and have been used in several important proof-of-principle experiments \cite{ruf2021quantum,bersin2024telecom}. Despite their success, the NV and group-IV defects both face a set of challenges, currently limiting their scalability. While the NV$^-$ offers excellent spin coherence, even at room temperature, and easy microwave-based ground state spin control \cite{togan2010quantum,childress2013diamond}, its susceptibility to spectral diffusion, due to a lack of symmetry \cite{faraon2012coupling,siyushev2013optically}, and a low Debye-Waller (DW) factor \cite{riedel2017deterministic} constitute major disadvantages. In contrast, the SiV$^-$ offers a DW factor $>$\, 0.7 and limited spectral diffusion, due to the center's inversion symmetry \cite{hepp2014electronic}. However, its orbital doublet ground state, split predominantly by spin-orbit coupling (SOC), results in phonon-induced decoherence of the $S=1/2$ ground state spin, limiting coherence times to the nanosecond regime, even at liquid helium temperatures \cite{pingault2014all}, requiring dilution refrigeration for longer spin coherence \cite{becker2018all,sukachev2017silicon}. The SiV’s heavier homologues GeV and SnV overcome this as they retain the SiV’s symmetry and associated favourable properties but offer stronger SOC and thus larger ground state splitting, resulting in lower thermal occupation of phonon modes at relevant frequencies and thus longer coherence times at liquid helium temperatures \cite{trusheim2020transform}. Alas, the stronger SOC makes spin control of these defects highly challenging, strongly pinning spin quantization axes of ground and excited states, making recent demonstrations of optical and microwave spin control particularly impressive \cite{debroux2021quantum,karapatzakis2024microwave,rosenthal2023microwave}. Moreover, both GeV and SnV feature zero phonon lines (ZPLs) in the 600\,nm regime, which are challenging laser wavelengths, requiring complex and costly systems such as dye lasers or frequency mixing systems. 

\begin{figure*}
\centering
\includegraphics[scale=1]{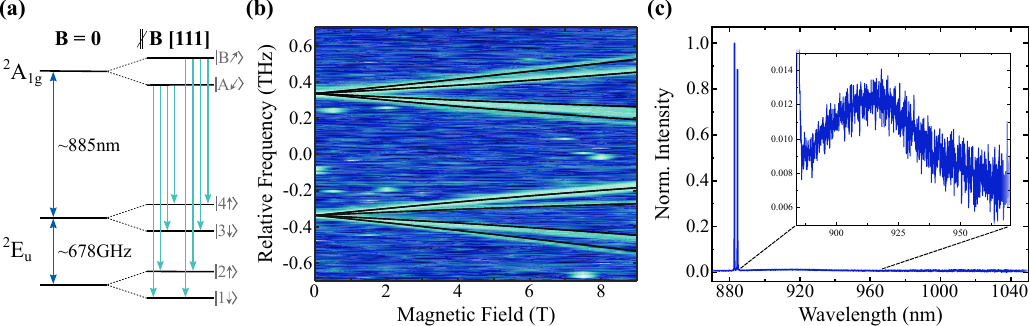}
\caption{\label{fig:elecstruc}\textbf{(a)} Theoretically predicted electronic structure of NiV$^-$ from \cite{thiering2021magneto}. \textbf{(b)} Experimental magneto-optical spectrum of NiV$^-$ at 1.6\,K. We used off-resonant excitation at 740\,nm at a power of 7\,mW before the objective lens. We integrated for 30\,s, repeated six times and then averaged with the magnetic field ramped up in 0.5\,T steps. We used the highest grating at 1800\,g/mm to achieve maximum resolution. The black lines  represent our group theoretical simulation. \textbf{(c)} Typical fluorescence spectrum for NiV$^-$ at 1.6\,K. Inset shows a zoomed-in view of the phonon sideband. We used 740\,nm excitation and 7\,mW of power before objective. The Debye Waller factor of 0.62 was determined by summing over the fraction of the spectrum of the ZPL from 882 to 886\,nm divided by the that light plus the light from the phonon sideband from 886\,nm to 1100\,nm. }
\end{figure*}

Recently, the negatively charged nickel vacancy (NiV$^-$) has been predicted to be a highly promising inversion-symmetric defect with potential to overcome some of these challenges \cite{thiering2021magneto}. Owing to its similar geometric structure, the NiV$^-$ ground state resembles that of the SnV$^-$, featuring a spin and orbital doublet ground state with strong SOC \cite{thiering2018ab}, resulting in a ground state splitting of approximately 670\,GHz (Fig.\,\ref{fig:elecstruc}(b)), enabling coherence times on the order of microseconds at readily achievable liquid helium temperatures. Specifically, approximating the coherence time as $\tau_{spin} = 1/\gamma_+$ where $\gamma_+$ is the acoustic phonon absorption rate, we find $\tau_{spin} = 0.12 \mu$s using the ground state splitting of the NiV and the parameters for group-IV vacancies from \cite{thiering2021magneto,iwasaki2017tin}. Furthermore, it provides two key advantages. First, the NiV$^-$ excited state features an orbital singlet and spin doublet, $S=1/2$, as a result of the Nickel 3d atomic orbitals mixing with the carbon dangling bonds \cite{larico2009electronic,thiering2021magneto} (see Fig.\,\ref{fig:elecstruc}(a) for proposed level structure), enabling efficient optical ground state spin control requiring only small transverse magnetic fields. This is because the excited state spin quantization axis will freely follow the applied magnetic field, while the ground state will be pinned to the $\langle 111 \rangle$ crystal axis by the strong SOC, allowing to drive previously spin-forbidden optical transitions. Secondly, it offers a much more convenient ZPL wavelength of 885\,nm, easily accessible with cheap and compact diode lasers. This wavelength can also be frequency converted to the telecom-c band around 1550\,nm in a single step with low added noise via difference frequency generation using readily available Thulium lasers at 2050\,nm enabling use of ultra-low-loss fibers necessary for a scalable quantum network \cite{bock2018high}. While telecom networking via upconversion of the SiV$^-$ has recently been demonstrated \cite{bersin2024telecom}, it is limited to the slightly higher loss telecom-O band due to Raman noise from the pump laser. Here, we report spectroscopic measurements consistent with its proposed electronic structure, find lifetime limited linewidths, and charge stabilize the defect.

We first perform cryogenic magneto-optical spectroscopy on NiV$^-$ centers created via ion implantation and high temperature high pressure annealing in a high temperature high pressure-grown sample to study the electronic structure of the defect \cite{supp}. Figure\,\ref{fig:elecstruc}(b) shows the fluorescence response of the NiV$^-$ in Faraday configuration when a magnetic field is applied at an angle of 109.5\,$\deg$ with respect to the high symmetry axis of the defect. We observe the initial two line spectrum split into eight components, consistent with the level structure described above. We can simulate the optical emission spectra by using a group theoretical model similar to the one developed by Hepp et al. \cite{hepp2014electronic} for the SiV$^-$ and taking into account the different excited state structure \cite{supp}. The Hamiltonian includes spin-orbit coupling, Jahn-Teller, as well as orbital and spin Zeeman interaction terms in the ground state, while the excited state solely features the Zeeman terms. Moreover, we include a correction term due to second-order Jahn Teller effects in the ground state \cite{thiering2021magneto}. Using the \textit{ab initio} predicted values initially, and then adjusting the strength of the SOC and JT interactions slightly, we overlay the simulated spectra onto the experimental spectra in Fig.\,\ref{fig:elecstruc}(b) and find close agreement. From this simulation, we identify the SOC as being the dominant interaction term splitting the ground state with a strength of 672\,GHz and Jahn-Teller at 8\,GHz. 

\begin{figure*}
\centering
\includegraphics[width=\textwidth]{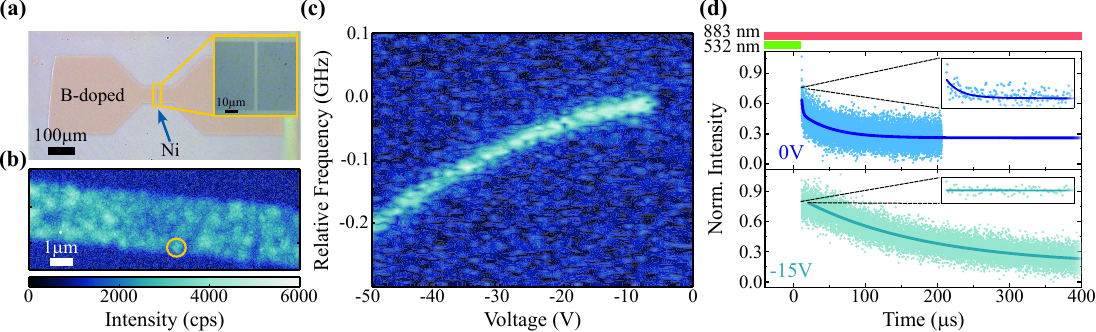}
\caption{\label{fig:pip} \textbf{(a)} Widefield microscope image of junction 1 with boron doped pads and nickel implantation region labeled. Top right insert shows zoom-in of junction. \textbf{(b)} Confocal microscope scan of junction at 1.6\,K with 692\,nm excitation at 1\,mW. Orange circle represents the location at which data was taken. \textbf{(c)} Revival of the NiV$^-$ fluorescence signal under resonant excitation by application of bias voltage across the p-i-p junction. The map displays many PLE scans at different bias voltages. The NiV$^-$ signal revives at $\sim$ -5\,V.  \textbf{(d)} Comparison of charge state lifetime by measuring the decay time of the NiV$^-$ signal after a 532 nm initialization pulse at 100 $\mu$W for 40 $\mu$s with constant resonant light being applied also at 100 $\mu$W. The top graph has no bias voltage applied while the bottom graph has -15V applied. Insets show the first two microseconds after the green pulse has finished. Intensity is normalized to the value when the 532 nm pulse is off, as compared with the IRF. }
\end{figure*}

Following the validation of the NiV$^-$ electronic structure, we then investigate the NiV$^-$ optical properties under off-resonant excitation and determine its DW factor. Figure\,\ref{fig:elecstruc}(d) shows a typical fluorescence spectrum with its two ZPL transitions arising from the spin orbit-split ground state and a weak phonon sideband. From this, we can calculate the DW factor as the fraction of light emitted into the ZPL to the total amount of light emitted by the defect. We find a DW factor of $\sim$ 62\,\%, comparable to the SnV$^-$ or SiV$^-$. This corresponds to a Huang-Rhys factor (average number of phonons emitted per excitation) of $S = -ln(DW)$ = 0.48 \cite{alkauskas2014first}.

\begin{figure}[b]
\includegraphics[scale=1]{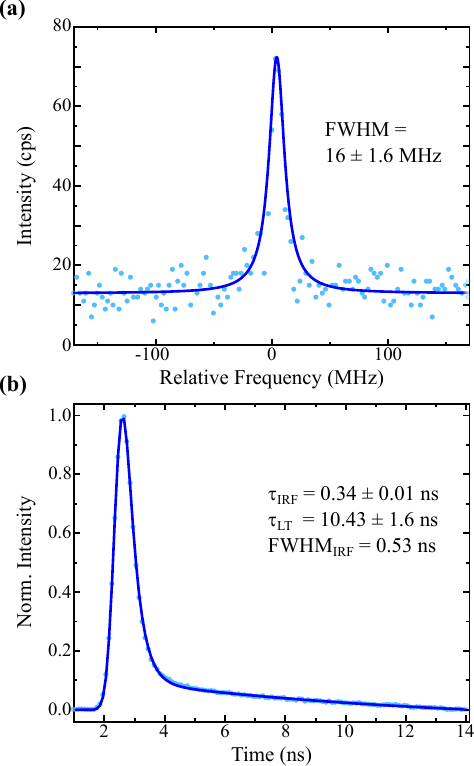}
\caption{\label{fig:tcspc} \textbf{(a)} Resonant (883\,nm) photoluminescence excitation scan across an NiV$^-$ optical transition at 0\,T and 1.6\,K with an applied bias voltage of -8\,V. The measured linewidth is 16\,MHz. Power before objective lens was measured at 30\,nW. \textbf{(b)} Time correlated single photon counting measurement to determine the excited state lifetime. Sample was excited using 698\,nm light at a power of 0.55\,mW through an NA=0.82 objective lens. We find an instrument response function (IRF) with a gaussian width of 0.53 ns. After deconvolution of the fast 0.34 $\pm$ 0.01 ns IRF, the excited state lifetime was found to be $\tau_{LT} = $10.43 $\pm$ 1.6\,ns. }
\end{figure}

Next, we characterize the optical properties under resonant excitation. Initially, when scanning the excitation laser across a ZPL transition, we observe complete termination of NiV$^-$ photo-luminescence (PL). Similar quenching upon resonant excitation has been observed in SnV centers \cite{gorlitz2022coherence} and is attributed to charge state instability. Moreover, we observe a sharp decline in PL when using off-resonant excitation above 700\,nm (see \cite{supp}), consistent with the predicted zero phonon line position of the neutral charge state NiV$^0$ [\cite{thiering2021magneto,iakoubovskii2000esr,lawson1993new}. While shorter wavelength light e.g. 532 nm can be used to revive the NiV$^-$ signal, it can also disturb the local charge environment which may negatively effect defect coherence properties and lead to spectral diffusion as has been observed for the SnV \cite{gorlitz2020spectroscopic,rugar2020narrow} and GeV \cite{chen2019optical}. Instead, to potentially more permanently stabilize the NiV$^-$ charge state or reduce the need for a short-wavelength optical re-pump, we fabricated all-diamond p-i-p junctions to gain direct control of the quasi-Fermi level (Fig.\,\ref{fig:pip}(a) and Fig. S1 of \cite{supp}), an approach which has proven successful for stabilizing the NV$^-$ and SnV$^-$ \cite{doi2014deterministic,luhmann2020charge}. In particular, two junctions each with two different $^{58}$Ni doses (1$\cross$10$^{12}$, 3$\cross$10$^{10}$, and 1$\cross$10$^{10}$, 3$\cross$10$^{9}$ ions/cm$^2$) were fabricated using a process similar to those fabricated for SnV \cite{luhmann2020charge} (details in \cite{supp}). By applying a bias voltage across the junction, we can tune the Fermi level to a regime of increased NiV$^-$ stability. Using this strategy, we demonstrate revival of NiV$^-$ PL signal (Fig.\,\ref{fig:pip}(c)) and identify lifetime-limited linewidths, necessary for producing indistinguishable photons.  

Confocal scans under off-resonant excitation at different bias voltages reveal a spatial dependence of the fluorescence across the junction similar to \cite{luhmann2020charge,de2021investigation,doi2014deterministic}, with the influence of the bias being most pronounced at the edges where the electric field is the strongest (see \cite{supp}). Hence, we select an emitter close to the junction edge in the most dilute implantation region ($3\cdot 10^9$ ions/cm$^2$) to perform resonant excitation experiments as shown in Figure\,\ref{fig:pip}(b). We here focus solely on the lower energy ground state linked to optical transitions around 883\,nm. Figure\,\ref{fig:pip}(c) shows a high resolution map created from photoluminescence scans at various bias voltages. The NiV$^-$ fluorescence remains quenched until a bias of roughly -5\,V is reached. For a semiconductor, the Fermi level is located close to mid-gap, making the NiV$^{2-}$ the favored charge state according to ab initio calculations in \cite{thiering2021magneto} (c.f.\,\cite{supp} for band bending calculations). By applying a bias voltage to the p-i-p junction, the Fermi level is pushed down, entering the NiV$^{-}$ charge state regime around -5\,V. Interestingly, although the p-i-p junctions are symmetric, we do not see revival of the NiV$^-$ when using positive voltages (reversed field direction). This in combination with the fact that the threshold voltage for NiV$^-$ revival coincides with the voltage at which current begins to pass through the junction (c.f.\,I-V curves in \cite{supp}) indicates that electron injection in addition to band bending may be involved.

We now discuss possible charge state conversion mechanisms which might explain the termination of PL under resonant excitation. 
Based on the ab-initio calculations by Thiering et al. \cite{thiering2021magneto}, it is reasonable to assume the charge conversion mechanism resembles those of SiV$^{-}$ \cite{rieger2024fast} and SnV$^{-}$ \cite{gorlitz2022coherence}, in which upon optical excitation of the NiV$^{-}$, an additional electron is optically excited from the valence band, filling the unoccupied ground state Kohn-Sham orbital and resulting in NiV$^{2-}$.
This is consistent with our observations in Fig.\,S4 of the Supplemental Material \cite{supp}, where NiV$^-$ fluorescence revival is observed under electrical bias on the positive (hole injection) side of the junction.

\begin{figure}[b]
\includegraphics[scale=1]{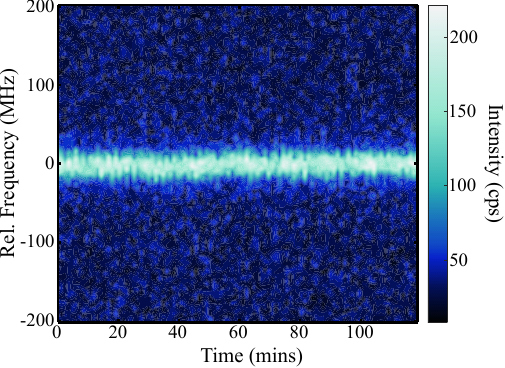}
\caption{\label{fig:linewidth} PLE map of a resonance over time using an applied bias voltage of -10V and an optical power before objective of 1\,$\mu$W were used. This was taken by scanning across the resonance repeatedly over time. }
\end{figure}

We probe the effect of bias on charge state lifetime by resonantly exciting the NiV$^{-}$, while applying 40 $\mu$s pulses of green (532 nm) light to initialize the NiV$^{-}$ charge state. After each pulse of green, we measure the decay of intensity as shown in Figure\,\ref{fig:pip}(d). With no bias voltage applied, we find both a fast decay time constant of 1.6 $\mu$s as well as a longer decay time constant of 42 $\mu$s. With a bias of -15\,V, we observe a slower decay. In fact, the intensity remains stable for about 15 $\mu$s after the green pulse ends, followed by a slow exponential decay with a time constant of 170 $\mu$s, indicating substantially improved charge state stability. Note that this decay occurs under continuous illumination. As such, the short optical pulses of ns to low $\mu$s length, as used for all-optical spin control \cite{debroux2021quantum}, are unlikely to induce charge transitions in biased centers.

Next, we explore the frequency shift of the transition as bias voltage increases. This effect is due to a DC Stark effect which can be modeled as
\begin{equation}
    \Delta E = -\Delta\mu F - \frac{1}{2}\Delta\alpha F^2, 
\end{equation}
with change in transition energy $\Delta E$, electric field $F$, and differences in ground and excited orbital state permanent dipole moment $\Delta\mu$ and polarizability $\Delta\alpha$, respectively \cite{harmin1982theory}. Using the Lorentz local field approximation $F = F_{ext}(\epsilon + 2)/3$ and diamond dielectric constant $\epsilon = 5.7$ \cite{spear1994synthetic} we convert between internal ($F$) and external ($F_{ext}$) electric fields. We calculate the latter with $F_{ext} = V_{bias}/d$ in which $d$ is the thickness of the junction which we find to be $\sim$3.5$\mu m$. With this, we can extract $\Delta\mu$ and $\Delta\alpha$ by fitting this model to the frequency-voltage map. We fit this model to data from 16 emitters and find an average $\Delta\mu = 6.9 \pm 4.7 \cdot10^{-4}$ GHz/MV/m which corresponds to $1.4 \pm 0.9 \cdot 10^{-4}$ Debye and $\Delta\alpha = -2.1 \pm 0.16 \cdot 10^{-4}$ GHz/$(MV/m)^2$ which corresponds to $2.47 \pm 0.1 \AA^3$.

This indicates that second-order effects dominate, as expected for an inversion symmetric defect with a vanishing permanent dipole moment. The values obtained are comparable to the SnV$^-$ \cite{de2021investigation}, and the permanent electric dipole moment is more than three-orders of magnitude less than that of the NV$^-$ \cite{tamarat2006stark,tamarat2008spin}. Along with the magneto-optical data, this provides further confirmation of the defect's assumed structure. Lastly, we find that the linewidths are relatively unchanged even with increasing bias, suggesting the use of electric fields for tuning the transitions over hundreds of MHz to enable overlapping of transitions for QIP applications.   

Lifetime-limited indistinguishable photons are a prerequisite for many photonic quantum networking and computing applications \cite{togan2010quantum,ruf2021quantum}. We characterize transition linewidths using photoluminescence excitation spectroscopy.  Across 12 emitters, we observe lifetime limited linewidths for 8 of them, with the remaining 4 showing slightly broader linewdiths of 24, 28, 19 and 18\,MHz, likely due to residual implantation damage. Figure\,\ref{fig:tcspc}(a) displays one such characteristic measurement. We measure the lifetime of the defect to be $\tau = 10.43 \pm 1.6$\,ns (Fig.\,\ref{fig:tcspc}(b)) using time-correlated single photon counting (TCSPC), in close agreement with a previous measurement of $\tau = 11$\,ns from \cite{orwa2010nickel}. This corresponds to a lifetime-limited linewidth of $\Delta\nu_{lim} = (2\pi\tau)^{-1} = 15.3 \pm 2.1$\,MHz. Thus across multiple emitters we observe linewidths consistent with the lifetime limit. From these emitters we estimate the width of the inhomogeneous distribution to be approximately 50\,GHz, with multiple emitters being within mutual Stark shift tuning range. Note that the NiV$^-$ is known to show improved spectral properties after HPHT treatment \cite{orwa2010nickel} which has not been performed after Ni implantation in this study to avoid potential damage to the p-i-p junctions. Further, with bias voltage applied, we find long term stability of these resonances (Fig.\,\ref{fig:linewidth}) with an RMS shift of the transition peak of 3\,MHz, likely limited by the long-term frequency resolution of our setup.

In conclusion, in this study, we have confirmed the NiV$^-$'s proposed inversion symmetric crystal structure as well as its electronic properties using magneto-optical spectroscopy in conjunction with a group-theoretical model. Furthermore, we show successful stabilization of the NiV$^-$ charge state under resonant excitation via electrical biasing in all-diamond p-i-p junctions, providing sufficient stability and a foundation for all-optical spin manipulation in the future. A second-order Stark shift provides further evidence for the defect's symmetry as well as a straightforward frequency tuning mechanism. Under electrical bias, we find stable transitions and linewidths consistent with the defect's measured lifetime limit. Our work provides a basis for future studies of NiV$^-$ spin properties and control and forms the foundation for its use as a qubit or spin-photon interface with properties overcoming challenges of current competitors such as the NV, SiV or SnV.

\begin{acknowledgments}
J.N.B is supported by the Cowen Family Endowment. I.A.M. acknowledges support from a Alfred J. and Ruth Zeits Endowed Fellowship from the MSU College of Natural Science. K.K. acknowledges support from an Lawrence W. Hantel Endowed Fellowship at MSU. This work was partially supported by the Fraunhofer USA Program Affiliate Cooperation for Knowledge Transfer, PACT.21.6. The authors want to thank A. Gali, G. Thiering, E. Poem, and B. Green for helpful discussions.
\end{acknowledgments}

\bibliography{bib}

\end{document}


\title{Supplementary Information: Lifetime-limited and tunable emission from charge-stabilized Nickel Vacancy Centers in Diamond}

\author{I.M. Morris}
\email{morrisia@msu.edu}
\affiliation{Michigan State University, Department of Physics and Astronomy, East Lansing, MI USA}

\author{T. Luhmann}
\affiliation{University of Leipzig, Felix Bloch Institute for Solid State Physics, Leipzig, Germany}

\author{K. Klink}
\affiliation{Michigan State University, Department of Physics and Astronomy, East Lansing, MI USA}

\author{L. Crooks}
\affiliation{Michigan State University, Department of Physics and Astronomy, East Lansing, MI USA}

\author{D. Hardeman}
\affiliation{Element Six Global Innovation Centre, Fermi Avenue, Harwell, Didcot, OC11 0QR, UK}

\author{D.J. Twitchen}
\affiliation{Element Six Global Innovation Centre, Fermi Avenue, Harwell, Didcot, OC11 0QR, UK}

\author{S. Pezzagna}
\affiliation{University of Leipzig, Felix Bloch Institute for Solid State Physics, Leipzig, Germany}

\author{J. Meijer}
\affiliation{University of Leipzig, Felix Bloch Institute for Solid State Physics, Leipzig, Germany}

\author{S.S. Nicley}
\affiliation{Michigan State University, Department of Physics and Astronomy, East Lansing, MI USA}
\affiliation{Coatings and Diamond Technologies Division, Center Midwest (CMW), Fraunhofer USA Inc., 1449 Engineering Research Ct., East Lansing, MI 48824, USA}

\author{J.N. Becker}
\affiliation{Michigan State University, Department of Physics and Astronomy, East Lansing, MI USA}
\affiliation{Coatings and Diamond Technologies Division, Center Midwest (CMW), Fraunhofer USA Inc., 1449 Engineering Research Ct., East Lansing, MI 48824, USA}

\maketitle

\begin{figure}[b]
\includegraphics[scale=0.75]{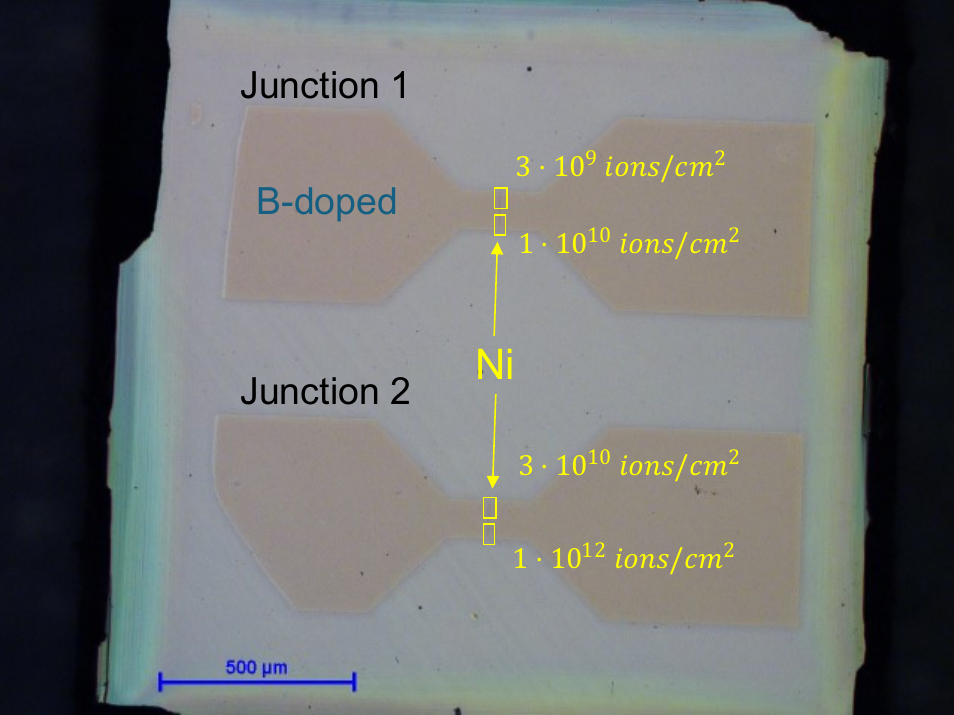}
\caption{\label{fig:micimg} Widefield reflection microscope image of the sample containing two p-i-p junctions with the boron-doped pads and nickel implantation regions labeled for both junctions. }
\end{figure}

\section{Sample Fabrication}
Two samples were used in this study. Both are single crystal diamond, but differ in type and fabrication method. For the magneto-optical spectroscopy and Debye-Waller factor measurements (Fig. 1 of main text), we used a (111)-oriented type IIa high pressure high temperature (HPHT) grown diamond substrate with 35(5) ppb of nitrogen measured with EPR, which was ion implanted with a fluence of $10^{12}$ Nickel ions/cm$^2$ $^{58}$Ni at 1MeV (performed at at RUBION, Bochum University, Germany). Following implantation, HPHT annealing was performed for two hours at 2000$^\circ$C and 8\,GPa (at Element Six). 

For the p-i-p device (see Figures 2, 3, 4 in main text), we used a (100)-oriented CVD grown electronic grade diamond from Element Six. Doping to create p-type regions was achieved via ion implantation of boron through a bow-tie shaped Ti-Au mask produced via e-beam lithography. Boron was implanted at two energies (40keV and 15keV) to obtain a constant doping level over a depth profile of more than 50\,nm with a thin ($<$40 nm) graphitic top layer. The sample was then thermally annealed at 900$^\circ$C for 1\,h followed by 3\,h at 1600$^\circ$C, all in vacuum to electrically “activate” the boron acceptors. The top defective layer with a vacancy density larger than $10^{22} \text{cm}^{-3}$ was turned to graphite. The sample was then cleaned for 4\,h in a boiling mixture of nitric, sulfuric, and perchloric acids to remove this graphite layer, leaving the highly p-doped layer at the surface. Nickel was implanted with an energy of 70keV and four different fluences ($3\times10^9$, $1\times10^{10}$, $3\times10^{10}$, and $1\times10^{12} \text{cm}^{-2}$) to compare the effects of bias on dense and dilute ensembles. An aperture of size 15 x 35 $\mu \text{m}^2$ was used to collimate the beam to the desired region within the intrinsic region of the p-i-p junctions. The sample was then annealed again at 1600$^\circ$C in vacuum followed by a soft oxygen plasma clean. 

\begin{figure}[b]
\includegraphics[scale=0.6]{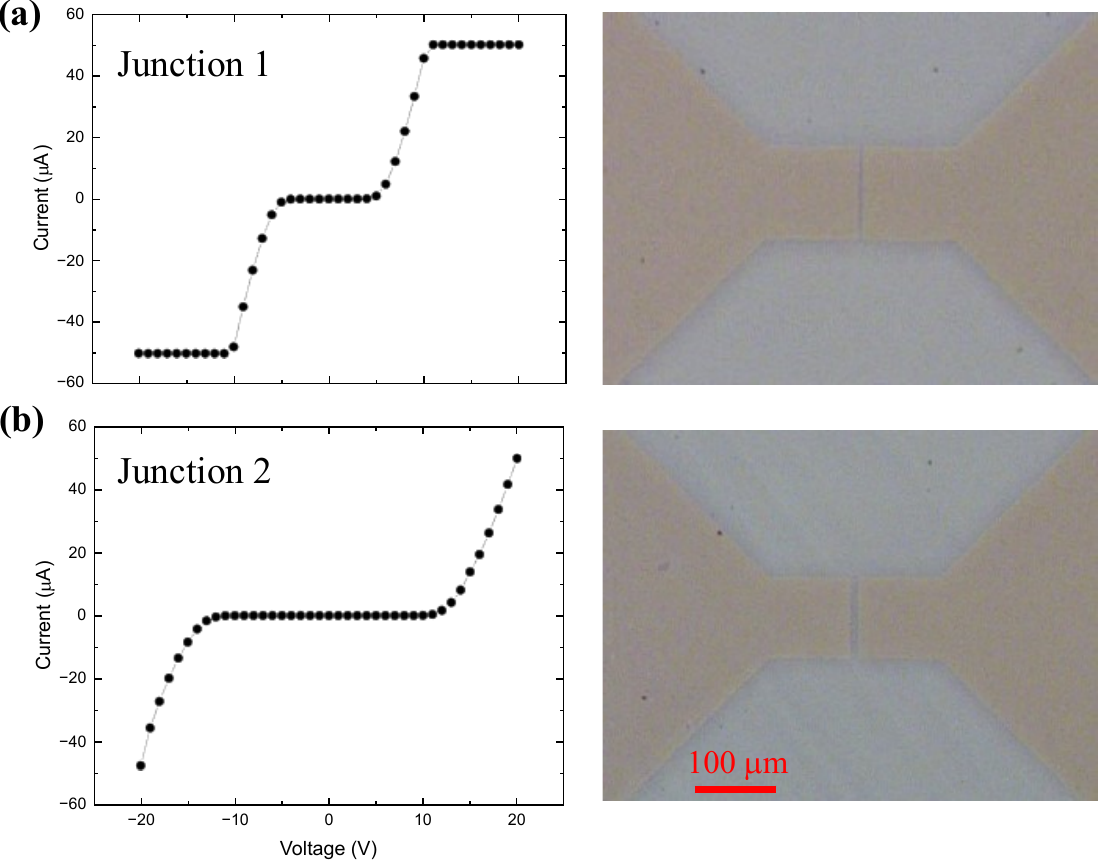}
\caption{\label{fig:iv} \textbf{(a)} Current-Voltage (IV) curves at room temperature on left and widefield image of junction 1 on right (junction width is 3.5\,$\mu$m). Voltage axis and scale bar from (b) are applied here. \textbf{(b)} Same format but for junction 2 (junction width is 5\,$\mu$m).  }
\end{figure}

\section{Experimental Setup}
To characterize the implanted NiV centers we placed the samples in a closed-cycle helium cryostat (attocube attoDRY 2100). Unless otherwise stated, all measurements were performed at 1.6K using a home-built confocal fluorescence microscope. A 100x cryogenic apochromatic objective lens (attocube LT-APO/VISIR) with NA=0.82 placed inside the cryostat was used for optical excitation and collection. The sample was corse positioned using a set of three piezoelectric slip-stick stages (attocube ANPX101, ANPz102). Confocal scanning was achieved using piezo scanners (attocube ANSxy100, ANSz100). The sample is located at the center of a superconducting solenoid coil which allowed for the application of a vertical (Faraday configuration) magnetic field of up to 9\,T. A Ti:Sapphire laser (Sirah Matisse CS) was used for both off-resonant and resonant experiments. A 532\,nm continuous wave diode-pumped solid-state laser (Specra-Physics Millennia eV 15) was used for optional optical revival of NiV$^-$ fluorescence under resonant excitation. . For off-resonant imaging, a 850\,nm long-pass filter (Thorlabs FELH0850) and a 880\,nm band-pass filter (Semrock FF01-880/11-25) were used in front of the detection fiber and single photon avalanche photodiodes (Excelitas SPCM-AQRH-16), respectively. For resonant excitation, the 880\,nm band-pass was used in excitation, and a 900\,nm long-pass (Thorlabs FELH0900) and 925\,nm band-pass (Edmund Optics 87-797, 50\,nm bandwidth) was used to detect on the weak phonon sideband. The Ti:Sapphire was then scanned across resonant transitions and the wavelength of the laser was read out using a wavelength meter (Highfinesse WS7-30). Fluorescence was transmitted via single mode fiber to either single photon avalanche photodiodes for counting or to a spectrometer (Teledyne Princeton Instruments HRS-750 with Blaze 400HRX) equipped with 600, 1200 and 1800 grooves/mm gratings. 

To determine the Debye Waller factor, spectra were taken by averaging four 20\,s exposures using the 600 grooves/mm grating. Spectra were background corrected and corrected for the spectral response function of the spectrometer and camera by normalizing to the manufacturer-provided wavelength-specific quantum efficiencies. 

For the p-i-p device, the junctions were electrically contacted using silver conductive paint and 38-gauge copper wires. IV curves were taken using a Keithley 2400 source meter. IV curves were taken at room temperature to confirm the Schottky nature of the junctions (see Fig. S2 and discussion below) before the sample was cooled to 1.6K.

\begin{figure}[t!]
    \centerline{\includegraphics[width=0.7\columnwidth]{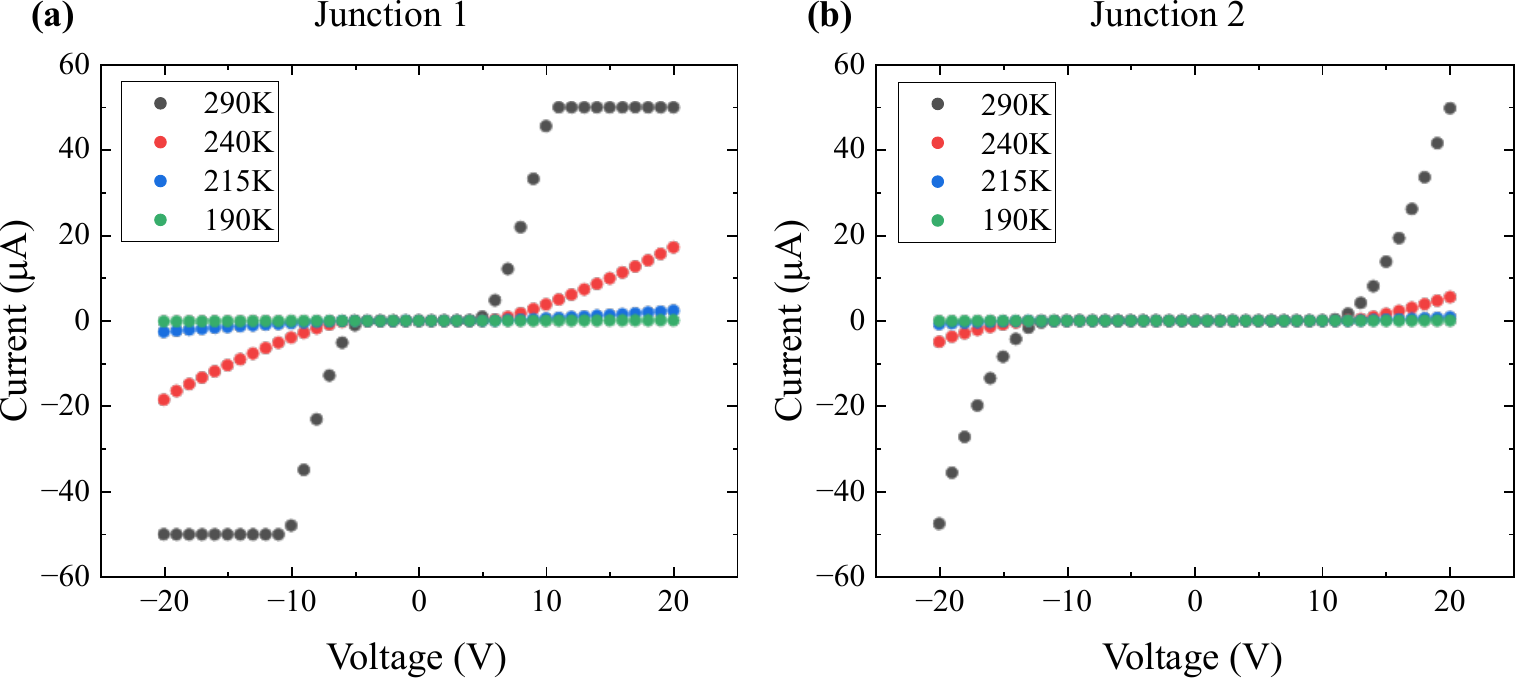}}
    \caption{\textbf{(a)} IV curves at varying temperatures for junction 1. \textbf{(b)} IV curves at varying temperatures for junction 2. }
    \label{fig:iv2}
\end{figure}

\section{Characterization of p-i-p junctions}
First, we took room temperature IV curves for both junctions as shown in Fig.\,\ref{fig:iv}. The current compliance limit was set to $\pm$50 $\mu$A, resulting in current plateaus in both graphs in Fig.\,\ref{fig:iv}. We note that the threshold voltage at which current begins to flow is different for both junctions, likely due to their difference in width with junction 1 having a width of $\sim$ 3.5 $\mu$m and junction 2 having a width of $\sim$ 5 $\mu$m. A temperature series of IV curves was measured while the sample was cooled to 1.6\,K, to determine the influence of temperature on the junction properties as seen in Fig.\,\ref{fig:iv2}. As expected, while the characteristic Schottky behavior of the p-i-p devices persists at lower temperatures, the measured current decreases with carrier density in the boron-doped regions. 

\begin{figure}[t!]
    \centerline{\includegraphics[width=1\columnwidth]{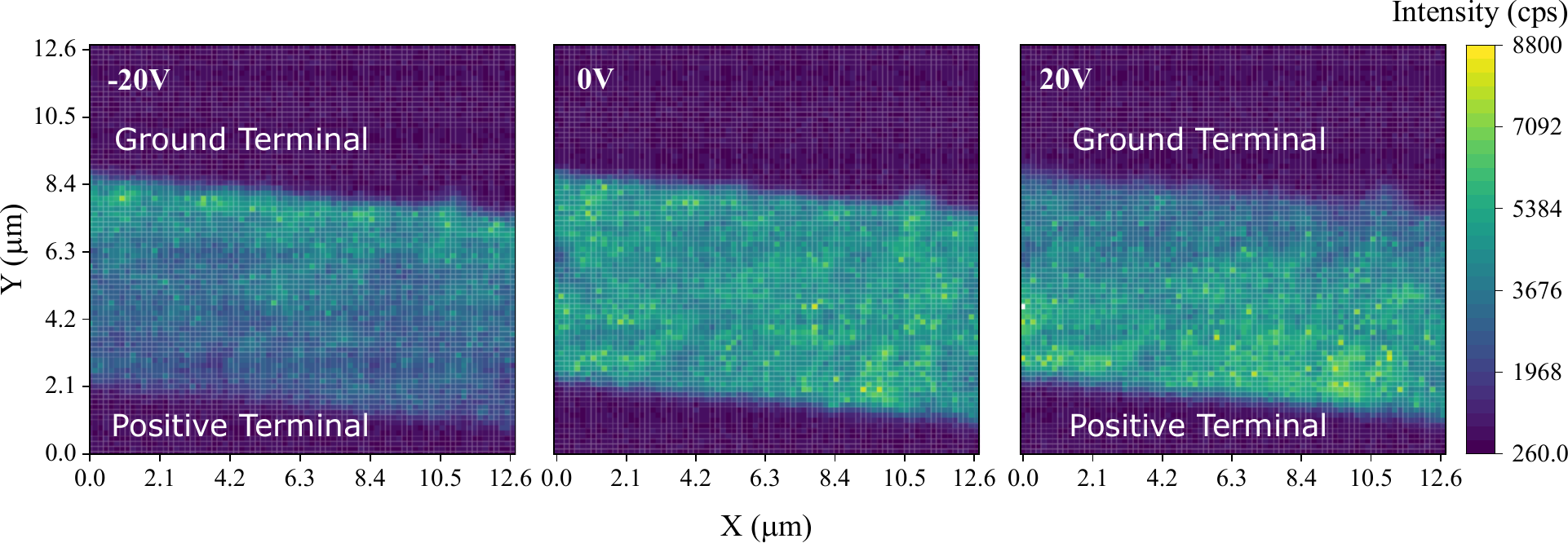}}
    \caption{Left to right, confocal microscope images of the 3$\cdot$10$^{10}$ ions/cm$^2$ region in junction 2 under different bias voltages.}
    \label{fig:confimg}
\end{figure}

Once the sample was cold (1.6K), we performed confocal scans of the $3\times10^{10}$ fluence region in junction 2 under different bias voltages to determine what if any effect bias had and its spacial dependence as shown in Fig.\,\ref{fig:confimg}. When the bias is applied, a clear increase of fluorescence along one edge and quenching along the opposite edge of the junction is observed, with the sides switching depending on the sign of the applied voltage. This has been observed for a similar p-i-p device used for NV and SnV centers \cite{luhmann2020charge} and is consistent with the band-bending calculations done below which show that the edges will have the most pronounced band-bending effect. Hence, we primarily study emitters along the edges of the junctions. 

Additionally, we conducted cross-correlation/hyperspectral imaging measurements by using two different filter windows for two avalanche photon detectors to check for the correlated revival/quenching of other defects when the NiV$^-$ fluorescence changes. In particular, we investigated the 700\,nm wavelength window. We tried this with both 690 nm excitation (filters: 700-40nm band pass and 700nm long pass) and 532 nm excitation (700-40nm band pass); however, we did not observe any cross-correlation nor spectral signal for the NiV$^0$.  
   
\section{Magneto-Optical Simulation}
The NiV$^-$ is predicted to be D$_{3d}$ symmetric like the group-IV vacancy complexes. As such, to simulate the magneto-optical spectra, we adapted the group theoretical model developed in \cite{hepp2014electronic,Hepp_2014} for the SiV, which has also been used to successfully reproduce spectra for the SnV \cite{trusheim2020transform} as well. The general form of all symmetry-adapted interaction Hamiltonians remains the same for the NiV$^-$, while changes were made to reflect the center's expected orbital singlet excited state. The model includes spin-orbit coupling, Jahn-Teller, as well as spin and orbital Zeeman interaction terms. Residual crystal strain and the effects of electric fields were neglected here. With that, we can express the ground state Hamiltonian in the basis $\ket{e_x \uparrow}$, $\ket{e_x \downarrow}$, $\ket{e_y \uparrow}$, $\ket{e_y \downarrow}$ and the excited state in the $\ket{a_{1g} \uparrow}$, $\ket{a_{1g} \downarrow}$ basis. The spin quantization axis is defined by the (111) high symmetry axis of the NiV which corresponds to the z-axis, with x and y orthogonal to it. Eigenvalues and eigenvectors were calculated in Matlab. The individual interaction Hamiltonians are listed below:\\

The spin-orbit interaction term is described by the following Hamiltonian \cite{hepp2014electronic,Hepp_2014}. 
\begin{equation}
    H_{SO} = \frac{\lambda}{2}\begin{pmatrix}
0 & 0 & -i & 0\\
0 & 0 & 0 & i\\
i & 0 & 0 & 0\\
0 & -i & 0 & 0
\end{pmatrix}
\end{equation}
where $\lambda$ is the spin-orbit coupling constant in the ground state.\\

The Jahn-Teller interaction term is described by:
\begin{equation}
    H_{JT} = \begin{pmatrix}
\xi_x & 0 & \xi_y & 0\\
0 & \xi_x & 0 & \xi_y\\
\xi_y & 0 & -\xi_x & 0\\
0 & \xi_y & 0 & -\xi_x
\end{pmatrix}
\end{equation}
where $\xi_{x/y}$ are the Jahn-Teller coupling strengths for the x and y components in the ground state.\\

These two interactions lift the two-fold degeneracy of the ground state by $\sqrt{\lambda^2 + 4\xi^2}$ where $\xi = \sqrt{\xi_x^2+\xi_y^2}$ is the strength of the Jahn-Teller coupling. The excited state has no orbital degeneracy and both of these interactions vanish accordingly.\\

For the Zeeman effect, the orbital component only includes the $L_z$ term (for detailed description, see Sec.\,2.2.6 from \cite{Hepp_2014}) and can be described by:
\begin{equation}
    H_{Zeeman,L} = p \gamma_L \begin{pmatrix}
0 & 0 & iB_z & 0\\
0 & 0 & 0 & iB_z\\
-iB_z & 0 & 0 & 0\\
0 & -iB_z & 0 & 0
\end{pmatrix}
\end{equation}
where p = 0.124 is the Ham vibronic reduction factor \cite{thiering2021magneto} and is largely responsible for quenching this orbital Zeeman component and $\gamma_L = \mu_B/\hbar$ is the orbital gyromagnetic ratio (where $\mu_B$ is the Bohr magneton and $\hbar$ is the reduced Planck constant). B$_z$ is the component of the external magnetic field along z. 

The spin component of the Zeeman interaction is 
\begin{equation}
    H_{Zeeman,S} = \gamma_S \begin{pmatrix}
B_z & B_x - iB_y & 0 & 0\\
B_x + iB_y & -B_z & 0 & 0\\
0 & 0 & B_z & B_x-iB_y\\
0 & 0 & B_x + iB_y & -B_z
\end{pmatrix}
\end{equation}
where $\gamma_S = 2\mu_B/\hbar$ is the gyromagnetic ratio for an electron and B$_x$ and B$_y$ are the components of the magnetic field along the x and y directions. 

Further, we added a correction term described in \cite{thiering2021magneto} which comes from a second-order Jahn-Teller effect and is described by:
\begin{equation}
    H_{2^{nd} Order JT} = 2\delta_pg_LS_zB_z
\end{equation}
where $\delta_p = 0.0839$ is the second-order parameter, $g_L = 0.7821$ is the Steven's reduction factor \cite{stevens1953magnetic}, and $S_z$ is the z-component of the Pauli spin matrices. 

Overall, the ground and excited state Hamiltonians are as follows:
\begin{equation}
    H_{gnd} = H_{SO} + H_{JT} + H_{2^{nd} JT} + H_{Zeeman,L} + H_{Zeeman,S} 
\end{equation}
and 
\begin{equation}
    H_{exc} = H_{Zeeman,S} 
\end{equation}

To optimize the model, we added inputs for the strength of the SO and JT interactions for the ground state, which could be adjusted to better fit the experimental data. For more details on the code implementation, see the appendix from \cite{Hepp_2014}.

\section{Stark Effect Measurements}
\begin{figure}[t!]
    \centerline{\includegraphics[width=1\columnwidth]{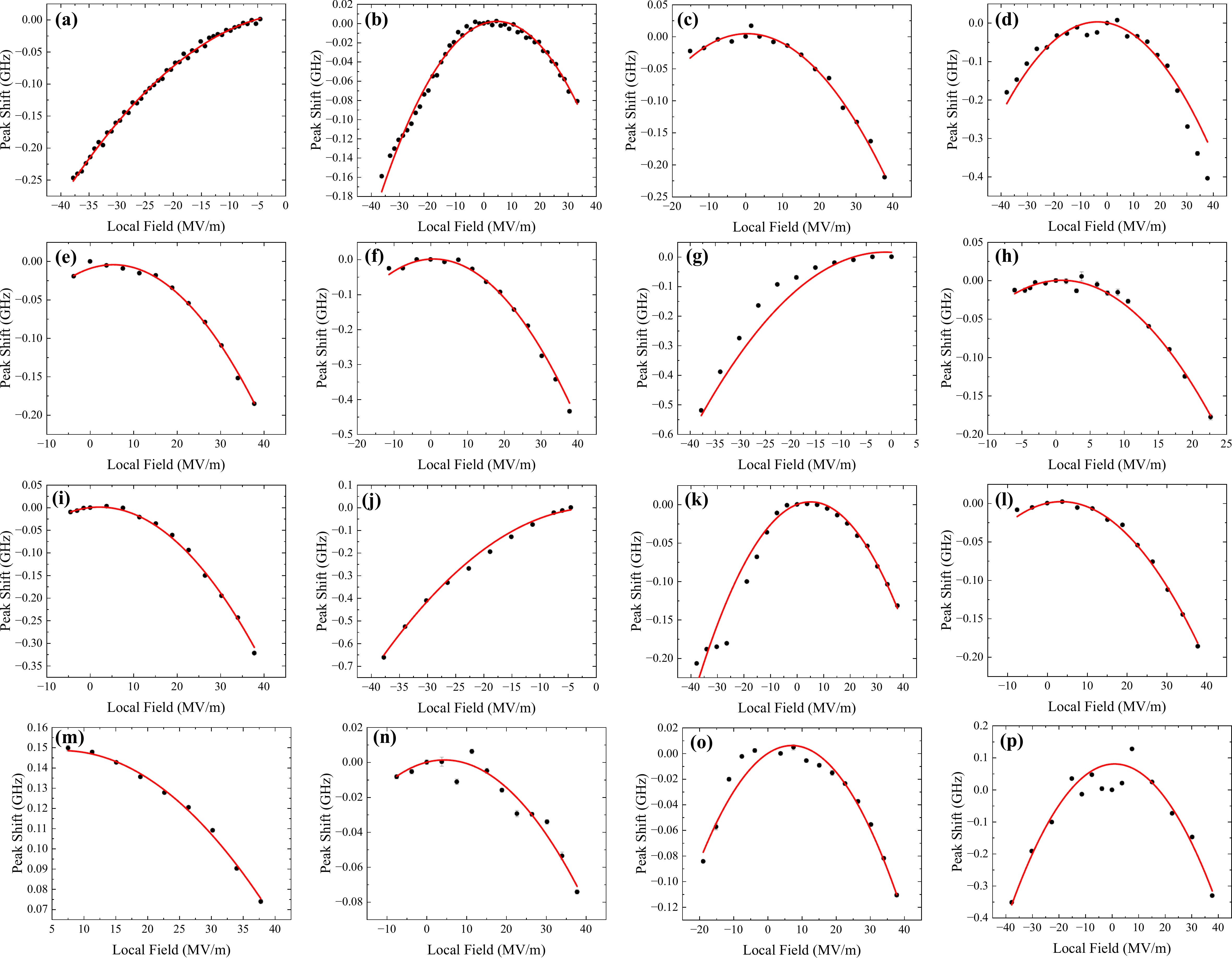}}
    \caption{\textbf{(a-p)} Stark shift of transition under resonant excitation as a function of the local field. (a) and and (j) were taken using solely resonant light, while the rest were taken using a green laser in conjunction with resonant light. }
    \label{fig:stark}
\end{figure}

A total of 16 emitters were investigated in terms of their electric field tuning behavior. Fig.\,\ref{fig:stark} shows extracted peak shifts as a function of bias voltage, which can then be fitted to a second order polynomial to extract the difference in dipole moment and polarizability as shown in Table 1. To extract the peak shifts, we performed PLE scans at different bias voltages, then fitted them to Lorentzians and subtracted the center frequency from either the center frequency at 0V or the center frequency at the lowest voltage at which the emitter was revived. The reason for the discrepancy between voltages measured is because of the voltage dependent nature for reviving the NiV$^-$ signal. We note that not all emitters revived under application of bias voltage alone and for these green light in conjunction with the resonant excitation was used. Interestingly, the use of green seems to largely open up the range of voltages which revive the NiV$^-$ signal, whereas without green, the emitters only revive at -5V and below. It should be mentioned that this is dependent on the side of the junction the emitter is on. For emitters on the left, -5V is the threshold while for emitters on the right, we found it to be flipped and a voltage of +5V is necessary. 

\begin{table}[b]
\caption{\label{tab:starkfit}
Extracted Stark shift values for $\Delta\mu$ and $\Delta\alpha$ along with the R$^2$ value for the fit. The first column letter corresponds to the labeled graph where the values were extracted from in Fig.\,\ref{fig:stark}. }
\begin{ruledtabular}
\begin{tabular}{cccccc}
 &$\Delta\mu$ (10$^{-3}$ Debye)&Error (10$^{-3}$ D)&$\Delta\alpha$ (\AA$^3$)&
 Error (\AA$^3$) & R$^2$\\
\hline
a& -0.18756 & 0.05402 & 1.88542 &0.04048
& 0.99696  \\
b& -0.19717 & 0.01251 & 1.25893 &0.01779
& 0.9799  \\
c& -0.02623 & 0.03497 & 1.88185 &0.03484
& 0.99502 \\
d& 0.26226 & 0.03809 & 2.17365 &0.07879
& 0.94244  \\
e& -0.36558 & 0.0417 & 2.04383 &0.04063
& 0.99529  \\
f& -0.09755 & 0.11148 & 3.59457 &0.15222
& 0.96841  \\
g& 0.19968 & 0.40071 & 4.92614 &0.39039
&  0.95573  \\
h& -0.13008 & 0.06592 & 4.44259 &0.13727
& 0.98857  \\
i& -0.18777 & 0.03459 & 2.88947 &0.04238
& 0.9968  \\
j& -0.32783 & 0.32783 & 5.00595 &0.23943
& 0.99794  \\
k& -0.27021 & 0.02446 & 1.46975 & 0.03391 & 0.97189\\
l& -0.22054 & 0.04093 & 1.86994 & 0.03997 & 0.9954\\
m& -0.03856 & 0.06358 & 0.70915 & 0.04488 & 0.99221\\
n& -0.12126 & 0.05364 & 0.77358 & 0.0508 & 0.95017\\
o& -0.35962 & 0.03783 & 1.53168 & 0.05351 & 0.94507\\
p& -0.11434 & 0.11714 & 3.48856 & 0.15728 & 0.91823\\
\end{tabular}
\end{ruledtabular}
\end{table}

\begin{figure}[t!]
    \centerline{\includegraphics[width=1\columnwidth]{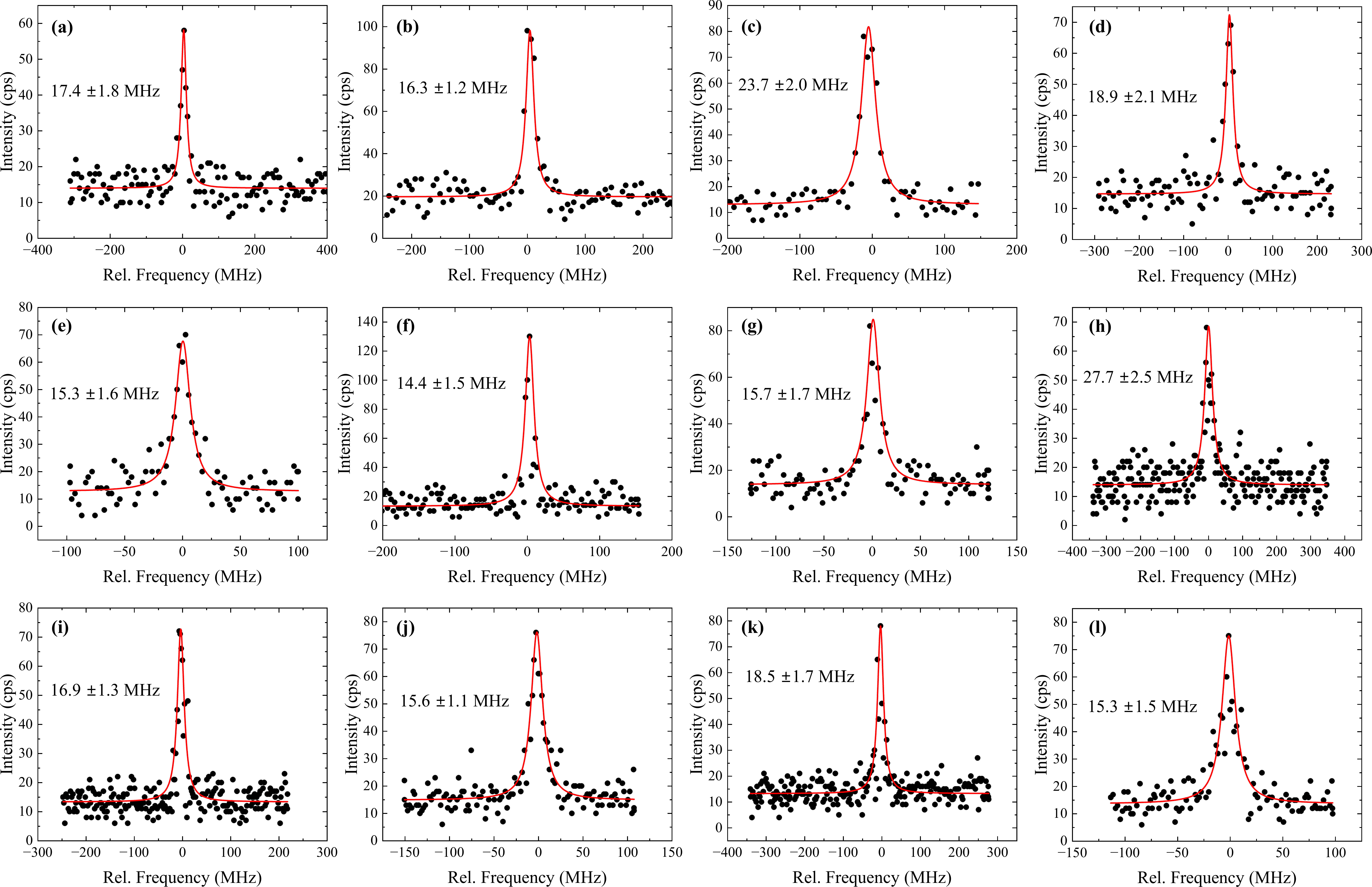}}
    \caption{\textbf{(a-l)} PLE scans across the optical transition at a variety of applied voltages for the spots measured, with the letter corresponding to the same spot at which the Stark shift graph data was taken in Fig.\,\ref{fig:stark}. The FWHM of the fitted Lorenztian is the number displayed within the graph.}
    \label{fig:ple}
\end{figure}

\begin{figure}[t!]
    \centerline{\includegraphics[width=0.8\columnwidth]{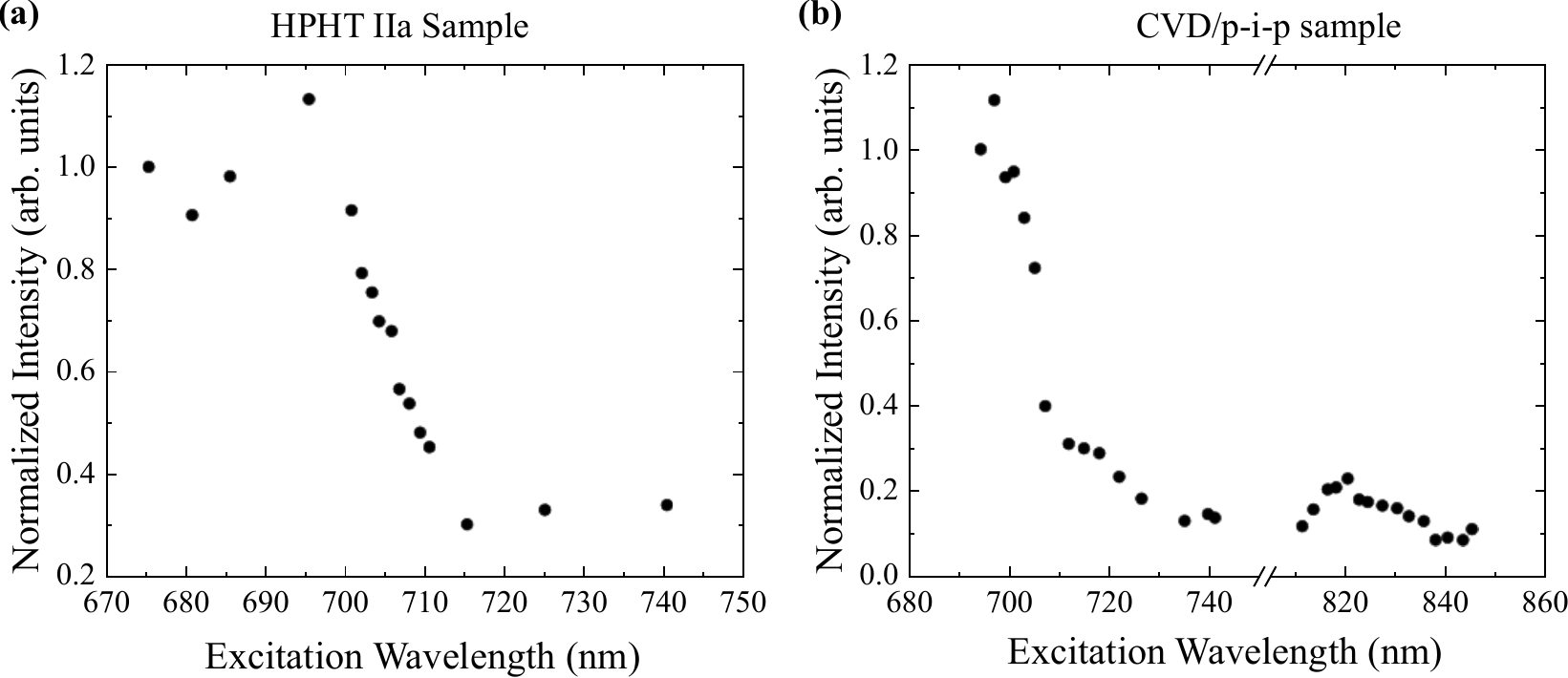}}
    \caption{NiV$^-$ intensity as a function of excitation wavelength. The power was kept constant across each wavelength used and the focus of the confocal was adjusted as the wavelength shifted. \textbf{(a)} was taken in the HPHT IIa sample, 1$\cdot$10$^{12}$ ions/cm$^2$ region. \textbf{(b)} was taken in the 3$\cdot$10$^{10}$ ions/cm$^2$ region of the p-i-p sample. The line break in \textbf{b)} is due to the first-order diamond Raman line overlapping with the NiV$^-$ detection window. }
    \label{fig:cpswav}
\end{figure}

\section{Charge State Stability Observations}

For band bending calculations of qualitatively identical p-i-p junctions we refer the reader to \cite{luhmann2020charge}. As can be seen from the charge transition levels calculated in \cite{thiering2021magneto}, the NiV$^{2-}$ is the dominant charge state for intrinsic diamond with the Fermi level at mid-gap. As bias voltage is applied to the junctions, the Fermi level is pushed down towards the NiV$^-$ regime.

We observe a sharp decline in photoluminescence (PL) when varying the excitation light from 690\,nm to around 710\,nm as shown in Fig.\,\ref{fig:cpswav}. Absorption data from \cite{lawson1993new} shows a localized absorption band in this region, similar to the ramp-up point in Fig.\,\ref{fig:cpswav}. Given that we expect the equilibrium charge state to be NiV$^{2-}$ (see above), we attribute this sharp change in excitation efficiency to the onset of the A$_{1g}$ bound exciton level which may non-resonantly decay into the 1.4\,eV excited state as predicted by Thiering et al \cite{thiering2021magneto}. Moreover, we tentatively attribute the local maximum around 820\,nm to excitation into a second excited state, similar to what was observed for SiV$^-$ \cite{haussler2017photoluminescence}.

Secondly, termination of PL under resonant excitation can be compared to recent studies for the SnV \cite{gorlitz2022coherence} and SiV \cite{rieger2024fast} defects, suggesting the dominant charge conversion process is the capture of an optically excited electron from the valence band by the negatively charged defect in its excited state, filling the unoccupied ground state orbital and resulting in the 2- charge state. Ionization e.g. via the conduction band or a nearby defect such as the divacancy (V$_2$) \cite{gorlitz2022coherence} is required to remove an electron to restore fluorescence. For both of these defects, their proximity to the valence band makes resonant photons sufficient to trigger this process. Similarly, the NiV$^-$ is also expected to be within 1\,eV of the valence band edge \cite{thiering2021magneto} and thus resonant light at 1.4\,eV would suffice to drive this process, too. In contrast, recovery via ionization to the conduction band is unlikely as it would require a >3 photon process under resonant excitation. The NiV$^{2-}$ is predicted to be excited by UV light and to emit at 2.96\,eV (419\,nm). Thiering et al. suggest that UV excitation can cause a charge transition from NiV$^{2-}$ to NiV$^{-}$ by ionizing an electron while also cause NiV$^{-}$ to NiV$^{2-}$ charge conversion by exciting an electron from the valence band, resulting in a dynamic equilibrium of both NiV$^{-}$ and NiV$^{2-}$ charge states \cite{thiering2021magneto}.  

In the p-i-p junction, we observe NiV$^-$ revival at around 5\,V. This voltage closely corresponds to the threshold voltage at which current starts to flow through the junction as seen from the IV curves, leading to electron (hole) injection in the inversion (depletion) regions along the edges of the junction. In combination with the fact that we see NiV$^-$ revival only on the hole injection side of the junction, this suggests that the injection of holes is an important part of reviving the NiV$^-$ in addition to band bending.

Similar effects were seen recently with SiV in \cite{rieger2024fast} in which SiV$^-$ PL was directly correlated with current flow. However, we note that we do not see such a direct correlation of PL intensity and current, potentially due to a saturation of intensity as we are dealing with a far less dense implantation of emitters.

In addition to charge state conversion only involving NiV, it is also possible for auxiliary defects to contribute to the charge state switching process. Here, we focus mainly on defects arising from the sample fabrication process itself as the electronic grade substrate only contains very low levels of impurities. In particular, we consider divacancies created during ion implantation as discussed by \cite{gorlitz2022coherence} as well as other nickel-related defects such as the nickel substitutional which will not anneal out even at high temperatures.\cite{pereira2002annealing}.

Using SRIM simulations \cite{ziegler2010srim}, we find 410 vacancies per Nickel ion are formed for 70\,keV implantation energy. Compared to what was found in \cite{gorlitz2022coherence}, this is nearly an order of magnitude less. However, we did not perform an HPHT treatment on the p-i-p sample which tends to reduce the concentration of residual concentration of vacancies and vacancy complexes. Thus, divacancies may still play a role in the re-ionization process of NiV$^{2-}$ and we tentatively attribute repumping using green light to a process like this.

Finally, we would like to note that the zero-phonon line of the neutral charge state, NiV$^0$ has been theoretically predicted to be around 700\,nm \cite{thiering2021magneto}. While being inconsistent with the expected charge state equilibrium and our observations in the pi-p junction, hypothetically, the sharp change in NiV$^{-}$ fluorescence with varying excitation wavelength around 700\,nm, could also be explained by a bi-directional charge transfer process involving NiV$^0$, whereby, at excitation below 700\,nm, it is excited to its $^3E_u$ excited state from which it may decay into the NiV$^-$ excited or ground state (state assignments taken from \cite{thiering2021magneto}). The inverse process could also occur whereby the NiV$^-$ $^2A_{1g}$ excited state decays into the excited state of the NiV$^0$. In contrast, at excitation wavelengths longer than the NiV$^0$ ZPL, only a unidirectional transfer from NiV$^-$ to NiV$^0$ can occur, without subsequent repopulation of NiV$^-$, somewhat similar to NV charge dynamics \cite{aslam2013photo,ji2018multiple} (see Figure 3b from \cite{aslam2013photo}). A competing process of NiV$^0$ to NiV$^-$ charge conversion at 1.22\,eV (1016\,nm) by addition of an electron from the valence band is predicted by Thiering et al. \cite{thiering2021magneto,londero2017identification}. In theory, at all wavelengths appropriate for NiV$^-$ excitation, this process occurs continuously, depleting NiV$^0$, further reducing the likelihood of NiV$^0$ involvement in the observed dynamics.

We suggest identifying cutoff wavelengths at energies higher than 532\,nm similar to what was done in \cite{gorlitz2022coherence} to determine if divacancies are a major contributor. This higher energy excitation may also cause ionization of the NiV$^{2-}$ charge state, which could further enhance the NiV$^{-}$.
  
\bibliography{supplementary_info}